\documentclass[twoside,twocolumn,9pt]{article}
\usepackage{extsizes}
\usepackage[super,sort&compress,comma]{natbib} 
\usepackage[version=3]{mhchem}
\usepackage[left=1.5cm, right=1.5cm, top=1.785cm, bottom=2.0cm]{geometry}
\usepackage{balance}
\usepackage{times,mathptmx}
\usepackage{sectsty}
\usepackage{graphicx} 
\usepackage{lastpage}
\usepackage[format=plain,justification=justified,singlelinecheck=false,font={stretch=1.125,small,sf},labelfont=bf,labelsep=space]{caption}
\usepackage{float}
\usepackage{fancyhdr}
\usepackage{fnpos}
\usepackage[english]{babel}
\usepackage{array}
\usepackage{droidsans}
\usepackage{charter}
\usepackage[T1]{fontenc}
\usepackage[usenames,dvipsnames]{xcolor}
\usepackage{setspace}
\usepackage[compact]{titlesec}

\usepackage{epstopdf}
\definecolor{cream}{RGB}{222,217,201}

\usepackage{hyperref}

\title{Betweenness Centrality as Predictor for Forces in Granular Packings}

\author{Jonathan E. Kollmer,$^{\ast}$\textit{$^{a}$} and Karen E. Daniels\textit{$^{a}$}}

\date{\today}

\begin{document}
\maketitle

\begin{abstract}
A load applied to a jammed frictional granular system will be localized into a network of force chains making inter-particle connections throughout the system. Because such systems are typically under-constrained, the observed force network is not unique to a given particle configuration, but instead varies upon repeated formation. In this paper, we examine the ensemble of force chain configurations created under repeated assembly in order to develop tools to statistically forecast the observed force network. In experiments on a  gently suspended 2D layer of photoelastic particles, we subject the assembly to hundreds of repeated cyclic compressions. As expected, we observe the non-unique nature of the force network, which differs for each compression cycle, by measuring all vector inter-particle contact forces using our open source PeGS software. We find that total pressure on each particle in the system correlates to its betweenness centrality value extracted from the geometric contact network. Thus, the mesoscale network structure is a key control on individual particle pressures.
\end{abstract}


\renewcommand*\rmdefault{bch}\normalfont\upshape
\rmfamily
\section*{}


\footnotetext{\textit{$^{a}$~Department of Physics, North Carolina State University, Raleigh, NC, USA. E-mail: jekollme@ncsu.edu}}

\footnotetext{\dag~Electronic Supplementary Information (ESI) available: [details of any supplementary information available should be included here]. See DOI: 10.1039/cXsm00000x/}



\section{Introduction}

For idealized granular particles in a jammed configuration, there are many different  ways for the force and torque balance on each particle to be satisfied for any given packing geometry and boundary conditions: this is known as the force network ensemble.\cite{Snoeijer2004,Tighe2010} A key reason for this ensemble of configurations is that particles deform at their contacts, and thereby store information about their loading history;\cite{Majmudar2005,Bililign2018} the rate- and asperity-dependence of real frictional contacts make this an even more prevalent effect.\cite{Dillavou2018}
Generally speaking, the inter-particle forces within these materials are mathematically under-determined: particle positions are insufficient to determine the force network. Furthermore, while two packings might have the same occupied volume or internal pressure, they can nonetheless have vastly different bulk material properties.\cite{Dagois-Bohy2012}
Therefore, a promising approach is to make predictions for the physical properties of granular materials using tools and concepts from statistical physics,\cite{Bi2015} by directly considering the ensemble of states. However, the choice of the correct ensemble remains the subject of active research.\cite{Bililign2018, blumenfeld-questions, tighe-gaussian, tighe-fluctuations, teitel-stress,sarkar}

\begin{figure*}
\centering
  \includegraphics[width=\textwidth]{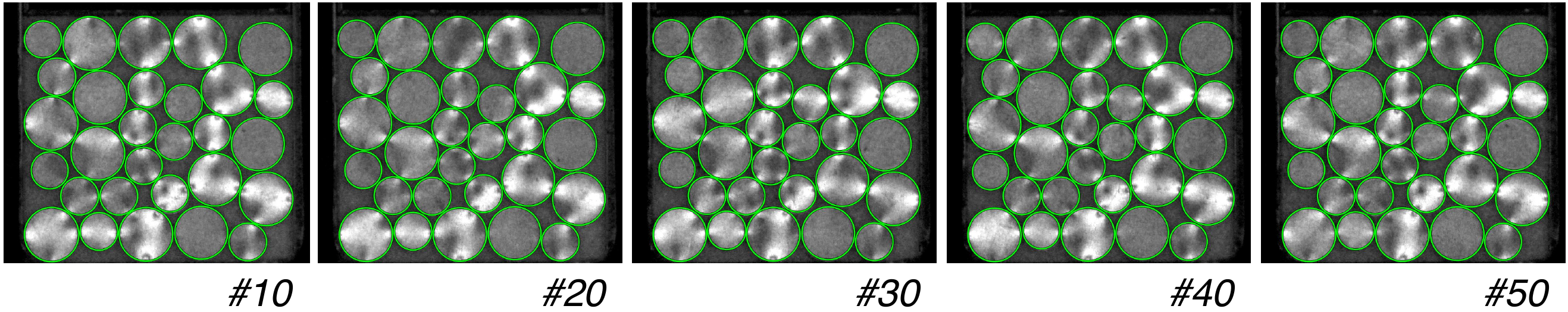}
  \caption{Five representative loading cycles for the same configuration of a granular packing, with $N=29$ photoelastic disks floated on an air table. In these darkfield images taken in polarized light, brighter particles are those with higher contact forces.\cite{Daniels2017} Particle outlines are superimposed as green circles.}
  \label{fgr:frames}
\end{figure*}

Importantly, the transmission of forces occurs primarily via linear structures running through a series of connected particles (see Fig.~\ref{fgr:frames}), forming a {\it force network}. This network provides the granular material with structures at many scales, from the particle (micro) scale to the force chain (meso) scale to the bulk (macro) scale. The mesoscale force network contains only a subset of the particles, and yet carries most of the load. Because these force chains can span significant portions of the system, it becomes inadequate to use either local or mean field approaches to describe the systems behavior.\cite{Cates1999} Thus, it is attractive to focus attention on force networks \cite{Coppersmith1996, Socolar2002, Peters2005, Owens2011} within jammed granular packings.

In order to make quantitative predictions about the role of force networks, it is useful to draw on the field of network science \cite{Newman2010, Bollobas1998, Smart2008, Bassett2015} in addition to statistical mechanics or grain-scale solid mechanics. Mathematical networks are often represented as a graph consisting of nodes and edges.\cite{Newman2010} In the case of a granular system, each node would represent a particle, and edges connect two particles (nodes) which share a contact. In many cases, it is informative to weight each edge by its contact forces; here we consider binary (unweighted) networks.  These techniques have recently been applied in a variety of granular contexts ranging from shear to compression to vibration \cite{Bassett2015, Huang2016, Papadopoulos2016, Pugnaloni2016, Kondic2016,Tordesillas2015, Herrera2011}
and a summary of the many applications can be found in a recent review.\cite{Papadopoulos2018}

As shown in Fig.~\ref{fgr:frames}, the force chain network that results from repeated experiments on the same particle network is both variable in the details, and repeatable in other aspects. For instance, the same two particles on the right side are often in strong contact with the wall. Thus, while the force-configuration fluctuates around several preferred states, some particles are loaded more often than others and the configurations are not completely random. This raises the question of what properties of the particle packing make it more or less likely for a particular particle to bear a strong load. 

This variability suggests that prior studies of the statistics of inter-particle contact forces \cite{Majmudar2005, Corwin2005,Howell1999} arise through simultaneously sampling two ensembles: the positions of the particles, and the valid configurations of forces. While there have been a few experiments probing granular ensembles,\cite{Puckett2013,Bililign2018} it has been difficult to decouple the external load that probes the force network from other forces acting upon the system, like gravity or basal friction.\cite{Kovalcinova2016} The experiments described here examine the distribution of forces in a loaded granular system, isolated from the effects of configuration.

\begin{figure}
\centering
  \includegraphics[width=0.8\linewidth,clip]{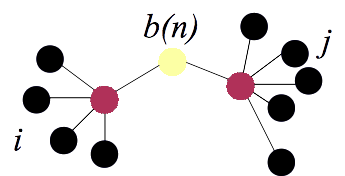}
  \caption{Schematic illustration of the concept of betweenness centrality, with each node color-coded by its $b$ value (brighter corresponds to larger values). The bright node at the center has high a high value of $b$ because many shortest-paths between any other two particles have to go through it. The dark nodes (here, arranged at the edges for clarity) have $b = 0$, since they never act as connecting nodes.
}
  \label{fgr:bc}
\end{figure}

In order to understand how the variability in force networks arises,  we focus on a mesoscale, nonlocal, measure of the network connectivity: the betweenness centrality $b$.\cite{Kintali2008}  As shown schematically in Fig.~\ref{fgr:bc}, the betweenness centrality of particle (node) $n$ is calculated by considering the extent to which shortest-paths between two other nodes must travel through that node. Mathematically, this is defined as the fraction of shortest path $s_{ij}$ along the edges (contact forces) between any two nodes $i \neq n \neq j$ in the system that goes through node (particle) $n$: 
\begin{equation}
b(n) = \sum_{i \neq n \neq j} = \frac{s_{ij}(n)}{s_{ij}}.
\label{eq:betweenness}
\end{equation}
In the results presented here, we performed our calculations using the open-source functions provided by the Brain Connectivity Toolbox.\cite{BCT}
Note that  $b$ depends only the adjacency network of the particles, rather than the inter-particle forces. This quantity, therefore, allows us to test how particle-scale forces depend on mesoscale connectedness. We find that $b$ can be used to predict the typical force on a particle, averaged over an ensemble of realizations.

\section{Experiment}

Our  experiment is designed to generate and characterize many different force configurations within a quasi-two-dimensional, hyperstatic (under-constrained)  granular packing at constant volume. To minimize extraneous external forces, we utilize a horizontal layer of photoelastic disks floating on a gentle air cushion.\cite{Puckett2012} This creates  an effectively gravity-free system without basal friction. The particles are laterally confined inside a piston that can apply an uniaxial load to the packing, via a series of quasi-static steps of constant wall-displacement.  

\begin{figure*}
\centering
  \includegraphics[width=0.7\linewidth,clip]{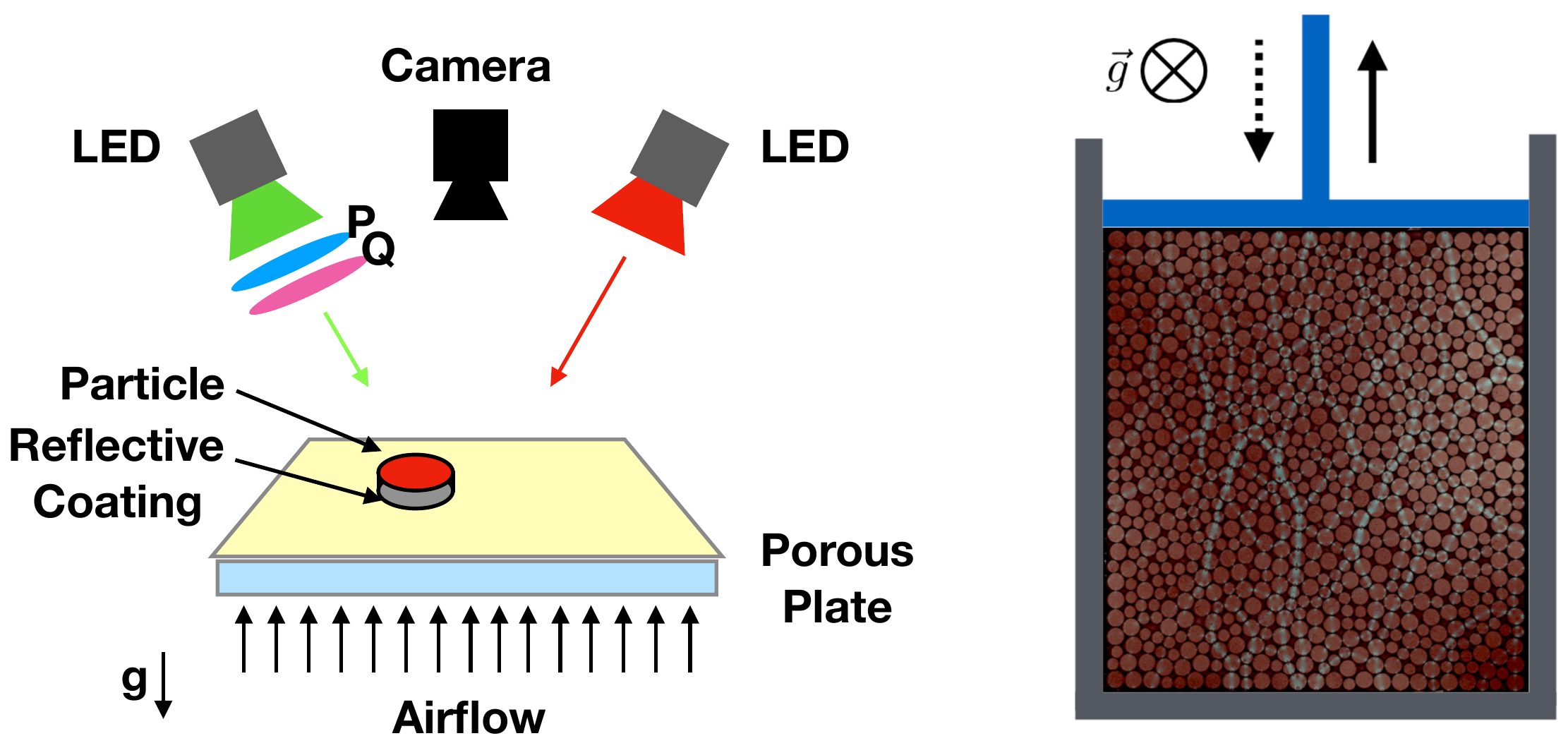}
  \caption{Experimental apparatus. 
Left: Photoelastic particles are floated on a horizontal air table to suppress the influence of gravity and basal friction,  and visualized from above. Illumination in the red channel is unpolarized, providing imaging of the particle positions. Illumination in the green channel is polarized, providing visual access to the internal stress field within each particle. The red and green color channels of the camera can later be separated to process particle positions and force information from a single color image.  Right: The grains are placed within a horizontal piston, and cyclically compressed. Particle positions are visible in red, and force chains in cyan.} 
  \label{fgr:setup}
\end{figure*}

A schematic drawing of the experimental setup is provided in Fig.~\ref{fgr:setup}. Each experimental run consists of cyclically loading and unloading the packing by a small enough wall displacement that (1) there is no change the nearest-neighbor particle configuration, but (2) the force network is erased between cycles. From cycle to cycle, there are microscopic changes in the contact points on the particles, but there are no neighbor changes. The initial volume $V_0$ was chosen to be just before the onset of jamming (zero pressure), and the final volume so that the mean contact force rises to approximately 0.5 N. This provides for both high resolution in force reconstruction, while also reducing the risk of out-of-plane buckling which occurs at higher forces. 
This maximum compression (minimum volume) was chosen to show a fully developed force network, while avoiding internal rearrangements or out-of-plane buckling. 
For the experiments presented here, we uniaxially compress the packing in steps of constant volume ($\Delta x = 0.2$~mm, corresponding to $\Delta V = 0.002869 V_0$). Each step is applied quasi-statically, with a waiting time between steps that is large enough to disregard inertial effects.

The granular material is composed of a bidisperse mixture of two different radii particles ($r_1 = 5.5$~mm and $r_2 = 7.6$~mm) to suppress crystallization. Between runs, the particle positions are randomized. We conduct experimental runs to generate an ensemble of force networks on two complementary system sizes: $N=29$ or $N=824$ particles. The $N=29$ particle system allows for hundreds of repetitions and high resolution imaging (which in turn allows for very precise force reconstruction), while the $N=824$ particle system allows for sampling an ensemble further away from the boundaries of the system.  The drawback of the larger system size, however, is that it is hard to re-compress the packing while also maintaining  the same particle configuration;\cite{Seguin2016} we find that only about one hundred cycles are typically possible. 

After starting an experiment, we discard the data for the initial $\approx 100$ cycles to get rid of the most prominent aging effects. After this initial annealing period, the force continues to fluctuate around a well-defined mean 
while the particle configuration remains unchanged.  For each compression cycle we take one image of the system, at a resolution of $2299 \times 2506$ pixels. 

In order to measure the vector contact forces on each particle, we utilize particles  made of a photoelastic material (Vishay PhotoStress PSM-4, elastic modulus $E=4$ MPa). Photoelastic materials have the property of rotating the polarization of transmitted light, by a known amount that depends on the local stress tensor. Therefore, it is possible to fit a theoretical model to images of the modulated light intensity within individual particles, and thereby obtain the vector contact force at each contact.\cite{Majmudar2005, Puckett2013, Daniels2017} We perform this step using our open source tool PeGS (Photoelastic Grain Solver) which is freely available on GitHub.\cite{PeGS} As a result, we get two vectorial contact forces for each contact in the system  ($\vec{F_{ab}}$ and $\vec{F_{ba}}$) since each side of a contact is processed individually. We then run PeGS again, giving the average of both forces as an initial guess. As a final result we then compare the residual of the force fit from both runs and pick the result with the smaller residual. This is to improve fitting accuracy and remove outliers. In order to measure the total pressure on each particle we calculate
\begin{equation}
P = \mathrm{Tr} ( {\hat \sigma }) 
\label{eq:pressure}
\end{equation}
where $ {\hat \sigma }$ is the stress tensor, derived from the photoelastic analysis.

In order to calculate network measures, we operate on the  binary contact (adjacency) matrix for each image. This is formed by identifying all contacts in the system for which we detect a nonzero value for the absolute force between two particles. We carefully checked that changing this threshold to $0.01$~N or $0.001$~N does not significantly change the results presented in this paper. At the end of the paper, we will additionally explore the successful use of a particle-separation criteria as an alternative method. 

\section{Results}

\begin{figure}
\centering
\includegraphics[width=0.49\linewidth,clip]{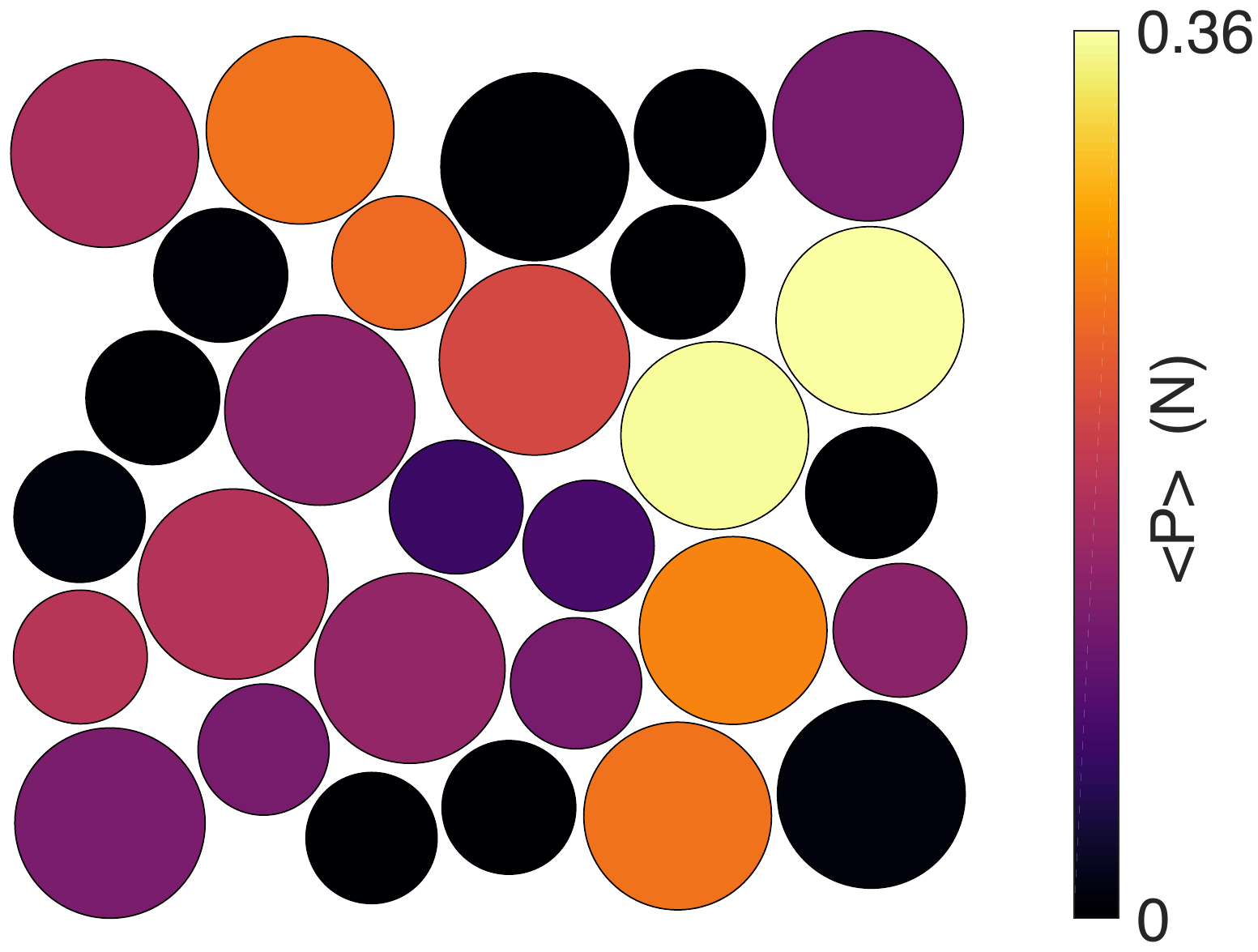}
\includegraphics[width=0.49\linewidth,clip]{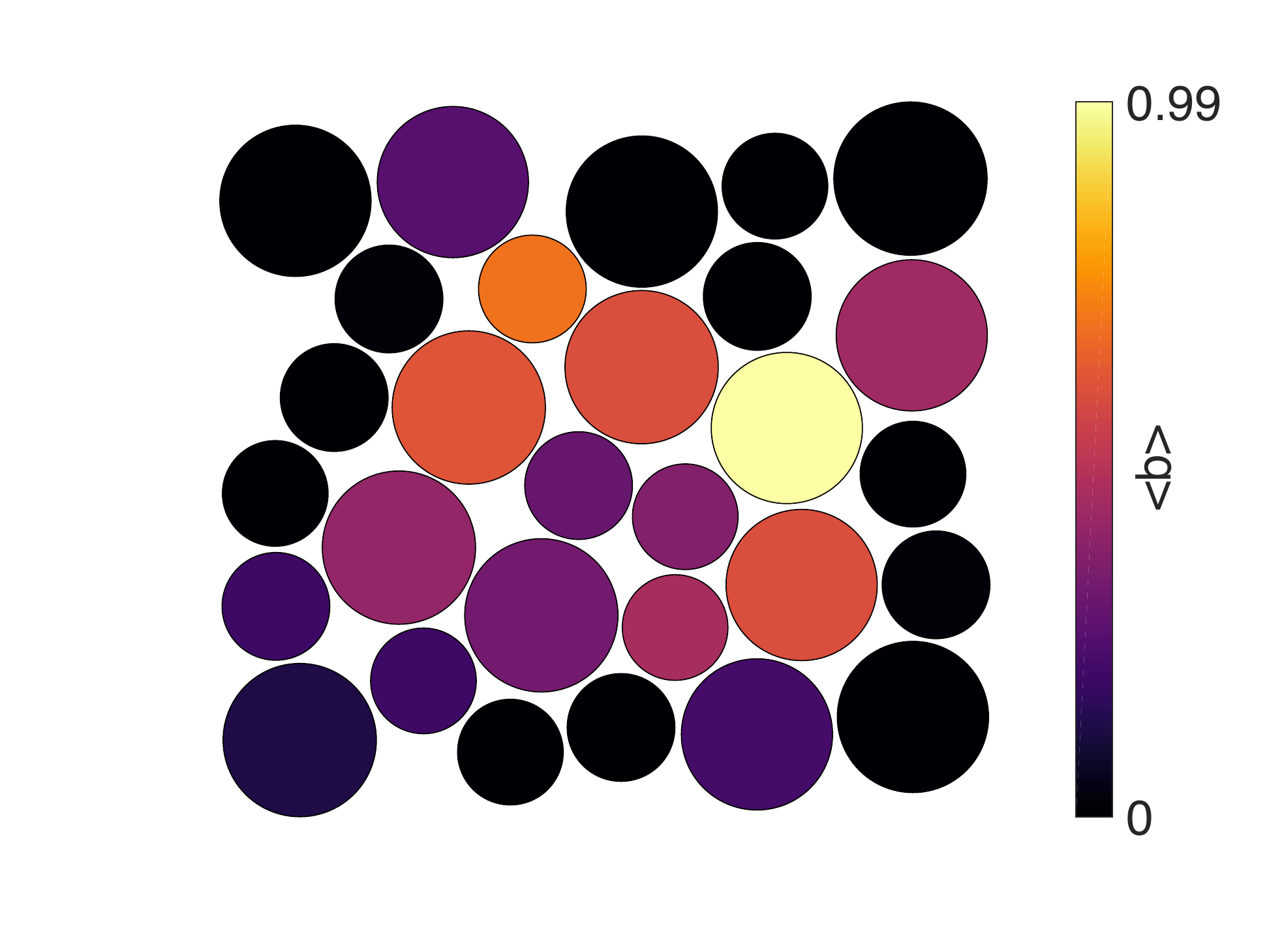}
\caption{Graphical representation of the average pressure  $\langle P \rangle$ (left) and the betweenness centrality $\langle b \rangle$ (right) on each particle in an  $N=29$ system. Averages are taken over 900 realizations of the configuration.}
\label{fig-29metrics}     
\end{figure}

\begin{figure*}
\centering
\includegraphics[width=\linewidth]{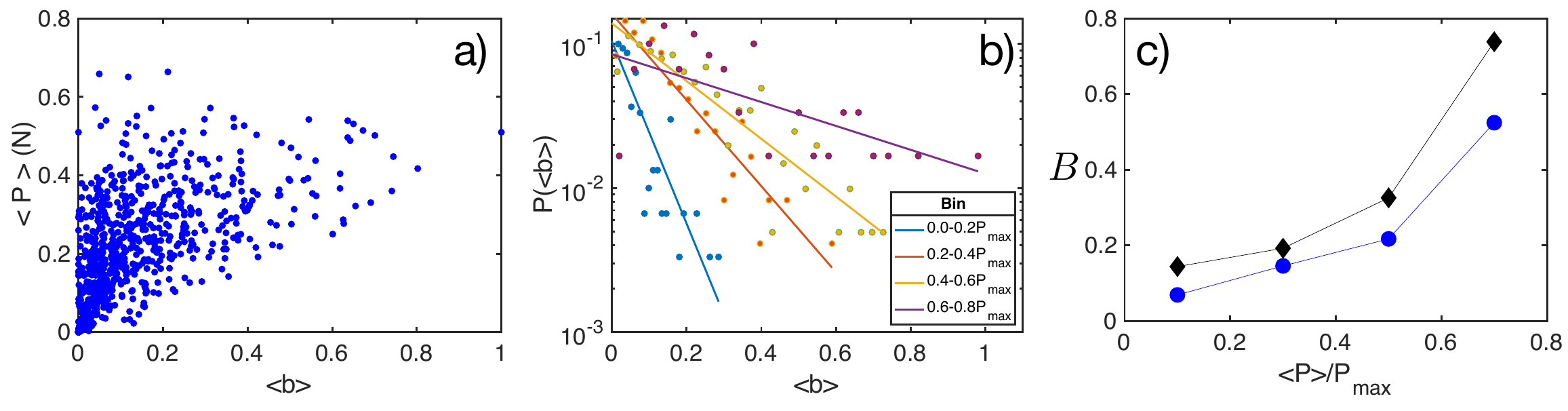}
\caption{
(a) Mean pressure $\langle P \rangle$ as a function of betweenness centrality $\langle b \rangle$, for each particle in an $N=824$ system. 
(b) Histogram of betweenness centrality values within four different pressure bins (symbols) from a), together with fits to Eq.~\ref{eq:exp} (lines). Each color represents a different bin. 
(c) Characteristic betweenness centrality value $B$, obtained from the fits shown in panel (b) (blue circles) and a second dataset (black diamonds) using the same system but a different particle configuration. 
Pressures were averaged over 88 (blue) and 48 (black) 
realizations of the same particle configuration.}
\label{fig-centralitypressure824}      
\end{figure*}

Figure~\ref{fig-29metrics} shows an analysis a typical run from the $N=29$ particle system illustrating the main result. In the left panel, each particle is colored by its mean pressure $\langle P \rangle$, averaged over the ensemble of 900 images in the run. Bright particles are those that are reliably on a strong force chain. In the right panel, each particle is colored by its average betweenness centrality value $\langle b \rangle$, obtained from the contact network. We take the average although the binary contact network barely fluctuates (i.e. the experiment was designed to avoid neighbor changes so most of the fluctuation is in magnitude of force). 
Visually, we observe the presence of correlations between the bright particles in both images. 

Next, we quantitatively explore this correlation using data from the $N=824$ particle system. Fig.~\ref{fig-centralitypressure824}a shows a scatter plot of $\langle b \rangle$ vs. $\langle P \rangle$ for each particle in the system, with averages  taken across 88 realizations (cycles) of the force network for a single particle-configuration. We quantify this positive correlation by finding a Pearson correlation coefficient $c_R = 0.60$  ($p = 1.3 \times  10^{-81}$). 
We repeat the experiment with a different configuration for 48 cycles and again get strong correlation results of $c_R = 0.47$  ($p = 6.1 \times  10^{-46}$). For the $N=29$ particle system the correlation yields $c_R = 0.75$  ($p = 2.3 \times  10^{-6}$). All three examined systems show a statistically-significant correlation.

To better visualize the underlying trend, we sort all data from a single experiment into four bins corresponding to four different pressure ranges. Fig.~\ref{fig-centralitypressure824}b shows a histogram of the corresponding measurements of betweenness centrality $\langle b \rangle$, taken within each of these four bins.
We observe that each of these four bins has a histogram following an exponential decay of the form
\begin{equation}
P \propto e^{- \langle b \rangle /B}
\label{eq:exp}
\end{equation}
where
$B$ is the characteristic value for that pressure range. As shown in Fig.~\ref{fig-centralitypressure824}c, this relationship is a monotonically increasing function, with the largest pressures (strongest, most reliable force chains) being particularly associated with large values of $\langle b \rangle$. 
Fig.~\ref{fig-centralitypressure824}c contains data from both datasets of the $N=824$ particle system. Data from the $N=29$ particle dataset (providing 29 values of $\langle b \rangle$) is too small to be suitable for this statistical evaluation.

The relationship between $\langle b \rangle$ and $\langle P \rangle$ establishes an underlying connection between the system's microscale (particle-scale pressure) and mesoscale (contact network) via the betweenness centrality parameter. This correlation makes physical sense by considering two very basic assumptions: that particles transmit force along inter-particle contacts, and that particles which have more such paths going through them will be more likely to accumulate forces from other particles. This relationship is only probabilistic, however, since  any given configuration of particles is (1) compatible with an ensemble of force networks \cite{Snoeijer2004,Tighe2010} and (2) contains forces which are history-dependent.\cite{josserand2000,paulsen2014, chaudhuri2010,bertrand2016,Majmudar2005}
This, while there is a clear positive correlation, there is also significant scatter in the data.

\begin{figure}
\centering
\includegraphics[width=0.8\linewidth,clip]{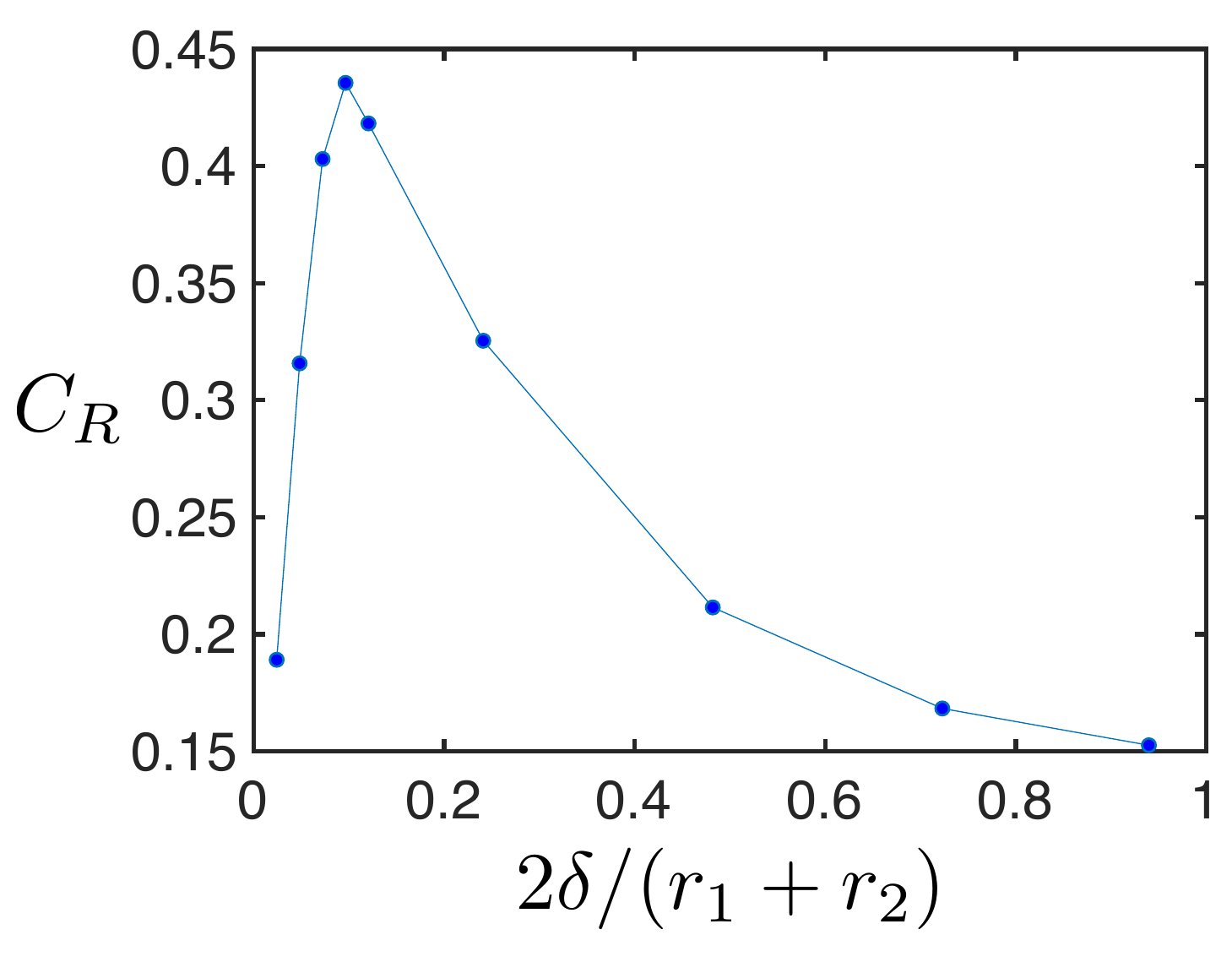}
\caption{Pearson correlation coefficient 
$c_R$ between the mean pressure $\langle P \rangle$ and the betweenness centrality $\langle b \rangle$, for each particle in the $N=824$ system, as a function of the particle separation threshold $\delta$. This is the same dataset as used in Fig.~\ref{fig-centralitypressure824}a, but for this analysis $\langle b \rangle$ is calculated from the geometric contact network rather than from thresholded forces.}
\label{fig-surfaceDistance}      
\end{figure}

Finally, we perform a stronger test of whether using thresholded force data to determine the contact network was pre-determining that we find this result. Using purely geometric data (rather than the photoelasticity of the particles), we determine the separation  $\delta = | {\vec x}_i - {\vec x}_j | - (r_i + r_j)$  between all pairs of adjacent $i,j$ particles. This gives an adjacency matrix that varies as a function of a threshold $\delta$, for which we can calculate $\langle b \rangle$ using Eq.~\ref{eq:betweenness}, and repeat the analysis shows in Fig.~\ref{fig-centralitypressure824}a for a range of different value for $\delta$. As shown in Fig.~\ref{fig-surfaceDistance}, our analysis still holds for a range around $\delta \approx 0.1 (r_1+r_2)/2$, 

which is consistent with the resolution to which we found the particle positions (approximately 1/10 of a particle radius). Therefore, it  is only necessary to identify particles that are close enough to be likely touching. The ability to perform an analysis of betweenness centrality without access to inter-particle forces opens the possibility to perform future studies on non-photoelastic particles. 

\section{Discussion and Conclusions}

We have designed an experiment that shows how a network measure known as betweenness centrality can predict the forces on a particle. Our predictions are tested using ensembles averaged over many realizations of the force network for a given particle configuration.  The betweenness centrality provides a new way to relate mesoscopic network features\cite{Giusti2016,Papadopoulos2016} to microscopic particle configurations. Therefore, its statistical moments might provide a simple tool for predicting such granular properties as the contact force distribution $P(f)$, heat transport coefficients,\cite{Vitelli2010} and the speed of sound.\cite{Makse1999,Bassett2012}

We have also showed that we can extract the necessary contact network information to make these predictions from purely geometric information, without the need for detailed photoelastic or other contact-force measurements. Whether or not force-network prediction can be done from only knowing  particle positions has been a recent issue of debate.\cite{Gendelman2016,DeGiuli2016} 
Here, we illustrate how to use experimentally-measured (imperfect) adjacency information to make probabilistic predictions for the inter-particle forces. We observe that this can be done without knowing the exact radii of each particle,\cite{DeGiuli2016} unlike the analytical solution of \citet{Gendelman2016} which is intractable under realistic circumstances. This insensitivity to contact measurements can be seen in Fig.~\ref{fig-surfaceDistance}, where even with 20\% error in radii measurements there is a strong correlation present. 

Importantly, we also observe variations in the force network among the  different samples in the ensemble (see (Fig.~\ref{fgr:frames}). Therefore, only statistical predictions make sense. However, the remaining variability is likely related to such phenomena as Branley's coherer, in which there is observed to be significant sensitivity of the electrical and thermal conductivity of metallic granular materials in response to electromagnetic waves.\cite{Falcon2005}

\section*{Acknowledgements}
We gratefully acknowledge James Puckett for the design and construction of the air table on which the apparatus is based, and for the inspiration for the new parallelized version of the contact-force code. We further acknowledge Estelle Berthier for helpful discussions. This research was supported by the James S. McDonnell Foundation and the NSF through grants DMR-0644743 and DMR-1206808.

\balance

\bibliography{ensembles} 
 \bibliographystyle{unsrt}

\end{document}